# Canalized hyperbolic magnetoexciton polaritons enabled by the Shubnikov–de Haas effect in van der Waals semiconductors

Guangyi Jia,[1,2,*] Qizhe Cai,[2] Chunqi Zheng,[2] Xiaoying Zhou,[3] and Cheng-Wei Qiu[2,†]

[1]*School of Science, Tianjin University of Commerce, Tianjin 300134, China*
[2]*Department of Electrical and Computer Engineering, National University of Singapore, Singapore, 117583, Singapore*
[3]*Hunan Provincial Key Laboratory of Intelligent Sensors and Advanced Sensor Materials, School of Physics and Electronics Science, Hunan University of Science and Technology, Xiangtan 411201, China*

Polariton canalization exhibits highly collimated and diffraction-free propagation characteristics in natural hyperbolic materials, holding great promise for molding the energy flow at nanoscale. Previously, the majority of canalizations are realized in phonon polaritons. Herein, we theoretically explore hyperbolic magnetoexciton polaritons (HMEPs) in van der Waals crystals of $WTe_2$, $MoS_2$, and phosphorene. Based on the Shubnikov-de Haas effect, canalized HMEPs with ultralow group velocity ($\sim 10^{-5}c$) and super-long lifetime (hundreds of microseconds) are predicted at low temperatures. We also show that the inter-Landau-level transitions and non-local dielectric screening effect significantly modify the optical topologies of canalized HMEPs, manifested by various exotic isofrequency contours (IFCs) including hyperbolic, witch-of-Agnesi, one-sheet, two-fold, and twisted pincerlike IFCs. Our findings reveal the significant effects of magneto-optical transport on hyperbolic polaritons, which not only enrich the interplay mechanism between magnetics and polaritonics, but also provide a promising playground for future developments in hyperbolic materials.

## I. INTRODUCTION

The development of twistronics and topological photonics has inspired great interest in the canalization of hyperbolic polaritons [1-5]. Polariton canalization is characterized by intrinsic energy-flow collimation and non-diffractive wave propagation, therefore, holds promise for various functional applications, e.g., super-resolution imaging, nanoscale energy transfer, and superlenses, etc [5-10]. It has been theoretically studied and realized in h-BN and α-$MoO_3$ and other hybridized systems [1-5,9-16]. Unfortunately, all of them are limited to phonon polaritons. The canalization of hyperbolic exciton polaritons (HEPs) has rarely been studied. The lack of relevant research may be due to the relative lag of discovering HEPs in two-dimensional (2D) perovskite crystals in 2018 [17]. Although seven years have passed, natural materials sustaining HEPs have been only expanded to a few van der Waals (vdWs) crystals including monolayer phosphorene [18] and black-arsenic [19], few-layer $WS_2$ and $WSe_2$ [20,21], tetradymites [22], and bulk CrSBr [23,24]. Besides, dispersions of HEPs are limited by intrinsic energy band structures in natural materials, which are unconducive to canalization controls of HEPs [18-25].

In fact, matter magnetization has been widely used to manipulate energy bands of materials [26,27]. 2D vdW crystals placed in an external magnetic field are characterized by a conductivity tensor $\hat{\sigma}(\omega)$=[$\sigma_{xx}$ $\sigma_{xy}$; $\sigma_{yx}$ $\sigma_{yy}$] with $\sigma_{xy}=-\sigma_{yx}$. The opposite-signed conductivity components (especially for imaginary parts) are just essential preconditions for achieving hyperbolic polaritons [28,29]. Additionally, an oscillation in the conductivity will occur at ultralow temperatures in the presence of an intense magnetic field, i.e., Shubnikov-de Haas (SdH) effect. SdH effect stems from Landau and Zeeman splitting level structure such that it can be used to rearrange energy bands in perfect monocrystals of practically all substances [30-32]. Especially, SdH effect at liquid-helium temperatures can excite extremely slow magnetoexcitons with lifetimes exceeding 1 ms [32-35], which are much more longevous than canalized phonon polaritons (only several picoseconds) [5,36,37].

One magnetoexciton is formed via an electron vacancy at a lower Landau level (LL) and an electron in some other state (orbital or spin) with a higher energy. Despite over 60 years of magnetoexciton study [38-40], previous works are mainly limited to magnetic-field-modulated LLs, diamagnetic coefficients, magnetoexciton condensate as well as related Faraday and Kerr effects [26,27,38-41]. Few of them are related to hyperbolic polaritons. Until recently, A. Y. Nikitin et al found that magnetized charge-neutral graphene ribbons support quantum hyperbolic magnetoexciton polaritons (HMEPs) at THz frequencies, and polariton canalization can be realized via suitably adjusting nanoribbon widths and array periods [42]. This work provides a novel type of actively tunable HEPs. Nevertheless, their designed hyperbolic structure still belongs to metamaterials whose structures need to be meticulously designed and precisely controlled through extremely high costs. Canalized HEPs have been scarcely studied in natural materials so far. Some completely open questions remain elusive. For example, how do optical factors including topology, group velocity, and lifetime of HMEPs evolute with LLs? How does the dielectric environment around magnetoexciton polaritons affect canalization and hyperbolic dispersion? Can HMEPs be canalized through pure magnetization rather than meticulously configurating geometric parameters of metastrutures?

In the present work, we focus on magnetoexcitons formed by interband transitions between LLs in valence and conduction bands, explore HMEPs resulting from the coupling between photons and anisotropic magnetoexcitons in natural 2D $WTe_2$, $MoS_2$, and phosphorene. We show that magneto-optical response has a rich impact on canalizations, dispersions, and optical topological transitions of HMEPs in the visible to near-infrared regimes. Importantly, HMEPs exhibit extremely slow relaxation ($\sim 10^{-5}c$, $c$ indicates light velocity in vacuum) and super-long lifetime up to hundreds of microseconds.

## II. MODEL AND THEORETICAL METHODS

In monolayer transition metal dichalcogenides 1T′-$MX_2$ (M= Mo, W; X=S, Te), the valence band mainly consists of $d$-orbitals of M atoms, while the conduction band mainly consists of $p_y$-orbitals of X atoms. The low energy $\boldsymbol{k}\cdot\boldsymbol{p}$ Hamiltonian for monolayer 1T′-$MX_2$ can be described by [43-47]

$$H_{k\cdot p}=\begin{pmatrix} E_c\left(k_x,k_y\right) & 0 & -i\gamma_1 k_x & \gamma_2 k_y \\ 0 & E_c\left(k_x,k_y\right) & \gamma_2 k_y & -i\gamma_1 k_x \\ i\gamma_1 k_x & \gamma_2 k_y & E_v\left(k_x,k_y\right) & 0 \\ \gamma_2 k_y & i\gamma_1 k_x & 0 & E_v\left(k_x,k_y\right) \end{pmatrix}, \tag{1}$$

where $E_c\left(k_x,k_y\right)=-\Delta-\alpha k_x^2-\beta k_y^2$ and $E_v\left(k_x,k_y\right)=\Delta+\lambda k_x^2+\eta k_y^2$ are the on-site energies of $p$ and $d$ orbitals in 1T′-$MX_2$, respectively, with $\Delta$ representing the $d$-$p$ band inversion at $\Gamma$ point in the Brillouin zone. The factors $\alpha$, $\beta$, $\lambda$, and $\eta$ are related to the reduced Planck's constant $\hbar$ and effective masses by $\alpha=\hbar^2/(2m_{cx})$, $\beta=\hbar^2/(2m_{cy})$, $\lambda=\hbar^2/(2m_{vx})$, and $\eta=\hbar^2/(2m_{vy})$, respectively. The parameters $\gamma_1=\hbar v_1$ and $\gamma_2=\hbar v_2$ with $v_1$ and $v_2$ denoting the velocities along the $x$ and $y$ directions, respectively.

For monolayer 1T′-$MoS_2$, these parameter values are $\Delta=-0.33$ eV, $\alpha$=7.62 eV·Å$^2$, $\beta$=23.8 eV·Å$^2$, $\lambda$=1.54 eV·Å$^2$, $\eta$=10.29 eV·Å$^2$, $\gamma_1$=2.55 eV·Å, and $\gamma_2$=0.30 eV·Å [46]. For monolayer 1T′-$WTe_2$, these parameter values are $\Delta=-0.489$ eV, $\alpha$=11.25 eV·Å$^2$, $\beta$=6.90 eV·Å$^2$, $\lambda$=0.27 eV·Å$^2$, $\eta$=1.08 eV·Å$^2$, $\gamma_1$=1.71 eV·Å, and $\gamma_2$=0.48 eV·Å [47]. For monolayer phosphorene, its Hamiltonian has a different form from Eq. (1), and its detailed calculations have been given in our previous work [43,44].

On the basis of a unitary transformation [46,47], the 4×4 Hamiltonian $H_{k\cdot p}$ in Eq. (1) can be separated into two decoupled 2×2 Hamiltonian blocks corresponding to "spin-up" and "spin-down" states, respectively. These two 2×2 blocks share the same dispersion relation. Thus, for brevity, we can only consider the spin-up Hamiltonian that reads [46,47]

$$H=\begin{pmatrix} E_c\left(k_x,k_y\right) & -iv_1P_x-v_2P_y \\ iv_1P_x-v_2P_y & E_v\left(k_x,k_y\right) \end{pmatrix}, \tag{2}$$

where $P_x=\hbar k_x$ and $P_y=\hbar k_y$.

Within the linear-response theory [43,44], the elements of conductivity tensor $\hat{\sigma}$ ($\omega$) for 2D materials in a perpendicular magnetic field $\boldsymbol{B}$ = (0, 0, $\boldsymbol{B}$) can be derived by

$$\sigma_{jk}=i\sigma_0\frac{\hbar^2}{l_B^2}\sum_{n',n,c,v}\frac{\left[f(E_{c,n'})-f(E_{v,n})\right]\langle c,n'|v_j|v,n\rangle\langle v,n|v_k|c,n'\rangle}{(E_{c,n'}-E_{v,n})(E_{c,n'}-E_{v,n}+\hbar\omega+i\Gamma)}, \tag{3}$$

where $j$, $k\in\{x, y\}$, $n$ (or $n'$) is the LL index of valence (or conduction) band, $\sigma_0=e^2/\pi\hbar$, $l_B^2=\hbar/(eB)$, $f(E_\xi)=\{\exp[(E_\zeta-E_F)/k_BT]+1\}^{-1}$ is the Fermi-Dirac distribution function with Boltzman constant $k_B$ and temperature $T$; $E_{v,n}$ and $E_{c,n'}$ are the LLs of valence and conduction bands, respectively; and $v_{j/k}$ are the components of group velocities. The sum runs over all states $|\zeta\rangle=|v,n\rangle$ and $|\zeta'\rangle=|c,n\rangle$ with $\zeta\neq\zeta'$. In this work, the level broadening factor $\Gamma$ and temperature $T$ are set to be 0.15 meV and 5 K, respectively. Besides, we take the Fermi energy $E_F$=0 such that the contribution from intraband transitions can be ignored [43,44,48,49]. Fig. S1 in [50] shows the LLs as a function of the magnetic field for the first 61 LLs in monolayer $WTe_2$. More detailed introductions on calculations of LLs and transition matrix elements $v_{j/k}$ can be found from Eqs. (S1) to (S12) in [50].

In the structural model, a sandwich structure, consisting of two 2D vdW crystals and an interlayer with a permittivity $\varepsilon_2$ and a thickness $d$, is deposited on a semi-infinite nonmagnetic substrate with a permittivity $\varepsilon_3$, as depicted in Fig. 1(a). Linearly polarized photons with an excitation energy $\hbar\omega$ impinge from air ($\varepsilon_1$=1) upon the sandwich structure, and the incident plane locates in $x$-$z$ plane. Unless otherwise specified, both upper and bottom 2D atomic crystals are 1T′-$WTe_2$ monolayers and their conductivity tensors are marked by $\hat{\sigma}_u$ and $\hat{\sigma}_b$, respectively. The conductivity component $\sigma_{xx}$ (or $\sigma_{yy}$) is along the armchair (or zigzag) orientation of $WTe_2$. The $\sigma_{xx,u}$ axis of upper $WTe_2$ has a rotation angle $\phi$ to the $x$-$z$ plane, and the $\sigma_{xx,b}$ axis of bottom $WTe_2$ has a twisted angle $\Delta\phi$ to the $\sigma_{xx,u}$. The structure is placed in a uniform magnetic field $\boldsymbol{B}=-B\hat{\boldsymbol{z}}$, and $B$=10 T by default unless otherwise indicated.

At tunning the rotation angle $\phi$ [see Fig. 1(a)], the anisotropy of monolayer atomic film is characterized via an effective conductivity tensor

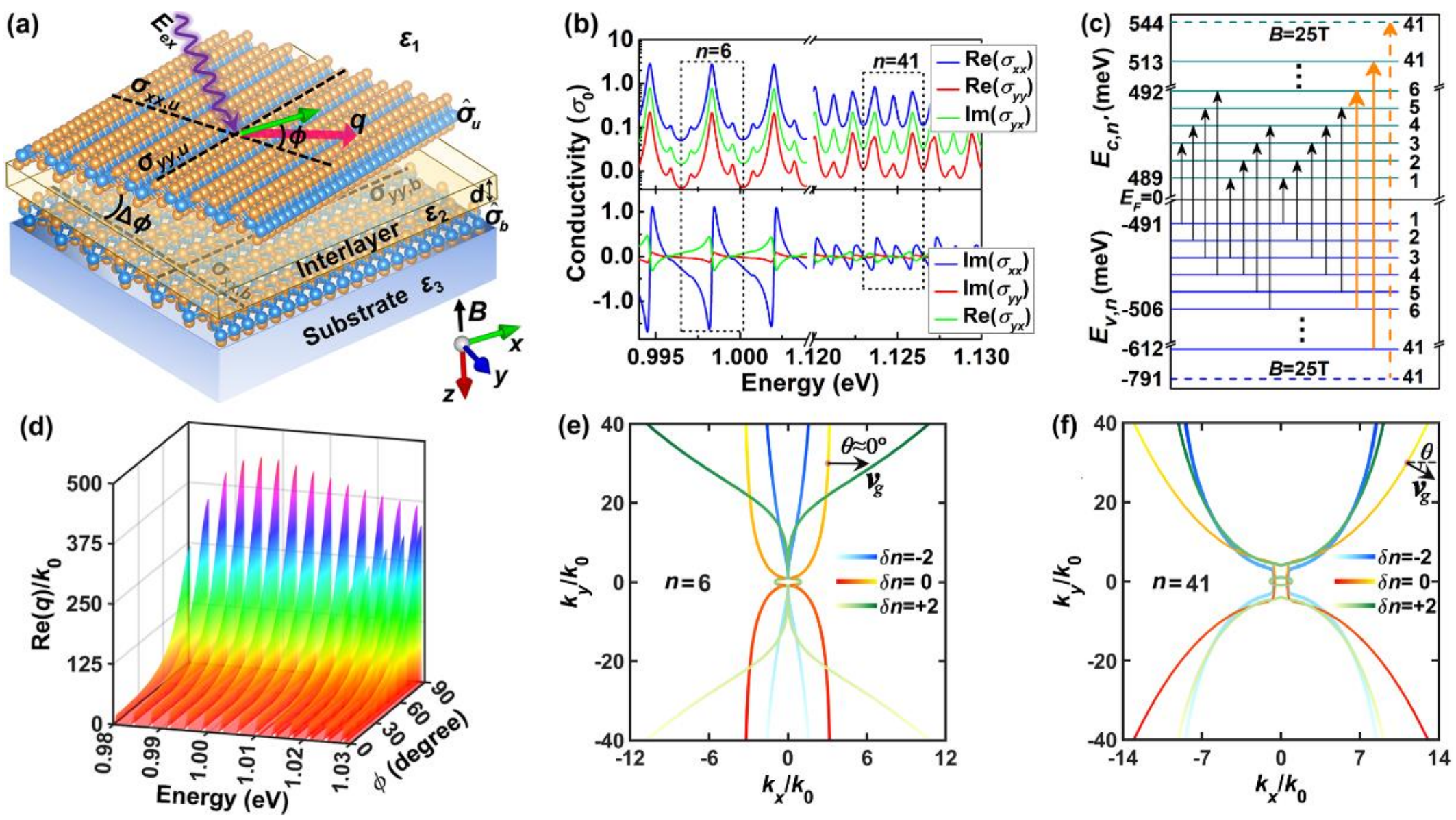

FIG. 1. (a) Schematic depiction of twisted bilayer vdW crystals in a static magnetic field $\boldsymbol{B}$. The wavy arrow marked by $E_{ex}$ indicates the excitation photons. The wavevector of surface waves $q=(k_x^2+k_y^2)^{1/2}$ is denoted by a pink arrow. (b) Optical conductivity spectra of monolayer $WTe_2$ versus the $\hbar\omega$ values. Oscillation periods at LL indexes $n$=6 and 41 are marked by two dotted-line boxes. (c) Schematic illustration of the interband magneto-optical transition rules in monolayer $WTe_2$, where numbers on right side indicate LL indexes (LLs $7\leq n(n')\leq 40$ are not sketched and are implied by ellipses), $|n=6\rangle\rightarrow|n'=6\rangle$ and $|n=41\rangle\rightarrow|n'=41\rangle$ are marked by orange arrows. At $B$=25 T, LLs $n=n'$=41 and $|n=41\rangle\rightarrow|n'=41\rangle$ are marked by dashed lines and an orange dashed arrow. (d) Azimuthal dispersions of surface waves in free-suspended $WTe_2$ versus the $\hbar\omega$ values. IFCs versus the transitions $\delta n$=0, ±2 of free-suspended $WTe_2$ at $n$=6 (e) and $n$=41 (f).

$$\hat{\sigma}_{eff}=\begin{pmatrix}\cos\phi & \sin(-\phi)\\ \sin\phi & \cos\phi\end{pmatrix}\begin{pmatrix}\sigma_{xx} & \sigma_{xy}\\ -\sigma_{xy} & \sigma_{yy}\end{pmatrix}\begin{pmatrix}\cos\phi & \sin\phi\\ \sin(-\phi) & \cos\phi\end{pmatrix}=\begin{pmatrix}\sigma_{pp} & \sigma_{ps}\\ \sigma_{sp} & \sigma_{ss}\end{pmatrix}, \tag{4}$$

where

$$\sigma_{pp}=\sigma_{xx}\cos^2\phi+\sigma_{yy}\sin^2\phi, \tag{5a}$$

$$\sigma_{ss}=\sigma_{xx}\sin^2\phi+\sigma_{yy}\cos^2\phi, \tag{5b}$$

$$\sigma_{ps}=(\sigma_{xx}-\sigma_{yy})\sin\phi\cos\phi+\sigma_{xy}, \tag{5c}$$

$$\sigma_{sp}=(\sigma_{xx}-\sigma_{yy})\sin\phi\cos\phi-\sigma_{xy}. \tag{5d}$$

Here, the nondiagonal response $\sigma_{ps}$ and $\sigma_{sp}$ can induce the couplings between $p$- and $s$-polarized waves. They arise not due to an intrinsic chirality of material, but from the magnetic-field-incurred Hall conductivity and a nonzero tilt of the incidence plane of light with respect to the main axis of 2D atomic material.

According to the generalized 4×4 transfer-matrix formalism [51], isofrequency contours (IFCs) of collective surface waves in the layered system in Fig. 1(a) can be deduced by the dispersion equation

$$\begin{aligned}&\left[\kappa_2^2\left(P_{12}^{++}P_{23}^{++}+P_{12}^{-+}P_{23}^{--}e^{-2\kappa_2 d}\right)+\varepsilon_2\sigma_{u,ps}\sigma_{b,sp}\left(1-e^{-2\kappa_2 d}\right)\right]\left[\varepsilon_2\left(S_{12}^{++}S_{23}^{++}+S_{12}^{-+}S_{23}^{--}e^{-2\kappa_2 d}\right)+\kappa_2^2\sigma_{u,sp}\sigma_{b,ps}\left(1-e^{-2\kappa_2 d}\right)\right]=\\&\left[\kappa_2^2\sigma_{u,sp}\left(P_{23}^{++}+P_{23}^{--}e^{-2\kappa_2 d}\right)+\varepsilon_2\sigma_{b,sp}\left(S_{12}^{++}-S_{12}^{-+}e^{-2\kappa_2 d}\right)\right]\left[\varepsilon_2\sigma_{u,ps}\left(S_{23}^{++}+S_{23}^{--}e^{-2\kappa_2 d}\right)+\kappa_2^2\sigma_{b,ps}\left(P_{12}^{++}-P_{12}^{-+}e^{-2\kappa_2 d}\right)\right]\end{aligned} \tag{6}$$

with

$$P_{12}^{\pm\pm}=\frac{\varepsilon_1}{\kappa_1}\pm\frac{\varepsilon_2}{\kappa_2}\pm i\sigma_{u,pp},\; P_{23}^{\pm\pm}=\frac{\varepsilon_2}{\kappa_2}\pm\frac{\varepsilon_3}{\kappa_3}\pm i\sigma_{b,pp},$$

$$S_{12}^{\pm\pm}=\kappa_1\pm\kappa_2\mp i\sigma_{u,ss},\; S_{23}^{\pm\pm}=\kappa_2\pm\kappa_3\mp i\sigma_{b,ss}.$$

In Eq. (6), all the conductivity tensor components are normalized to $c\varepsilon_0/4\pi$, $\kappa_{j=1,2,3}=[(qc/\omega)^2-\varepsilon_j]^{1/2}$ are inverse penetration depths of the surface waves into the upper and lower media. The $\kappa_j$ are normalized to $k_0=\omega/c$, and $q=(k_x^2+k_y^2)^{1/2}$ denotes the wavevector of surface waves.

If there is no the interlayer and the bottom 2D material, the multilayer structure in Fig. 1(a) will become the case that monolayer vdW semiconductor is deposited upon a semi-infinite substrate with a permittivity $\varepsilon_3$. Then Eq. (6) can be simplified to the following form

$$(\kappa_1+\kappa_3-i\sigma_{ss})(\varepsilon_1/\kappa_1+\varepsilon_3/\kappa_3+i\sigma_{pp})=\sigma_{ps}\sigma_{sp}. \tag{7}$$

If we further set the factors $\varepsilon_3=\varepsilon_1=1.0$, it will correspond to the condition of suspended monolayer vdW semiconductor in a free space.

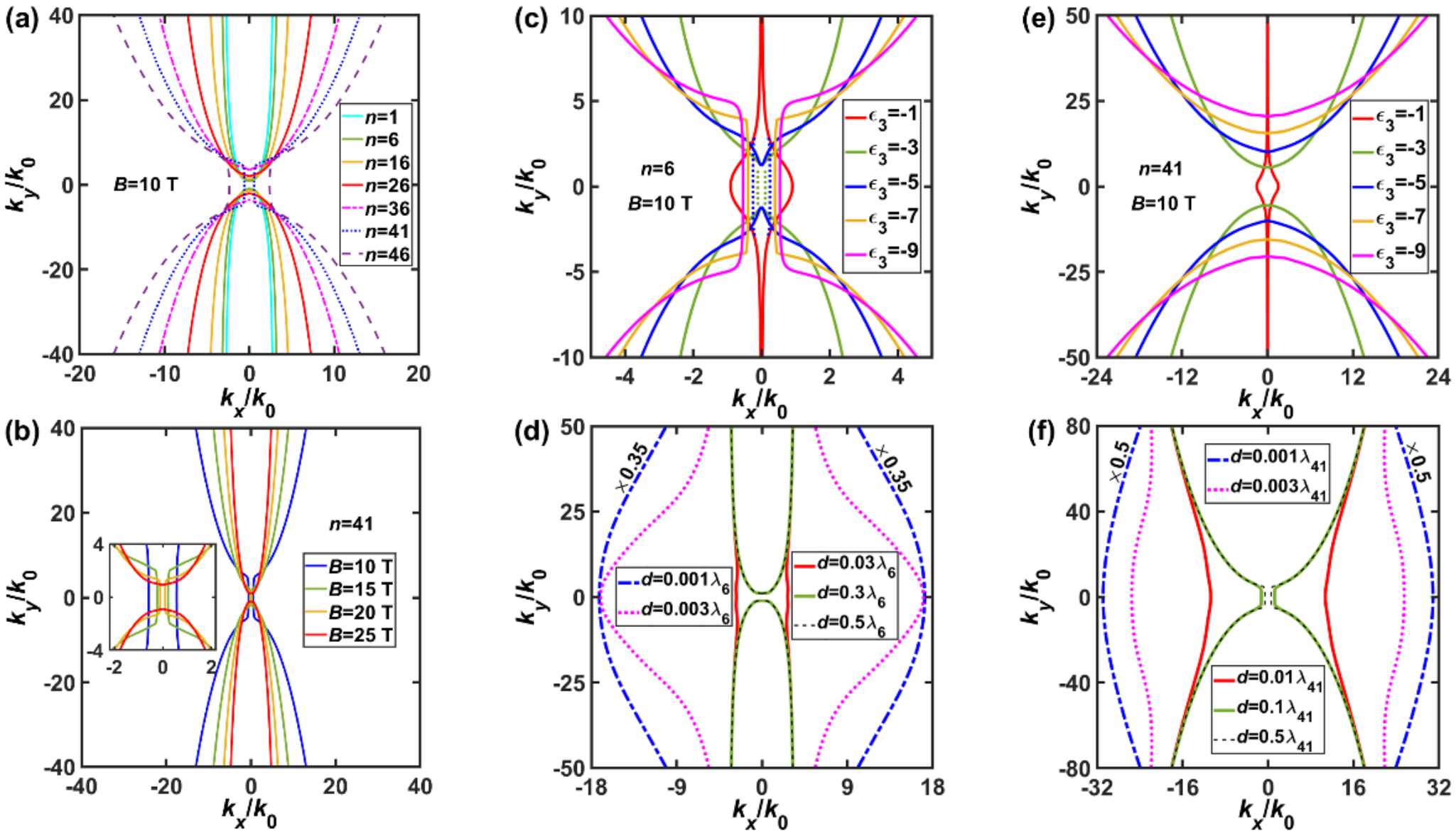

FIG. 2. (a) For free-suspended monolayer $WTe_2$ at $B$ = 10 T, variations of hyperbolic IFCs (stem from *p*-polarized wave) with respect to the LL index *n* of interband transitions $\delta n = n' - n = 0$. (b) Variations of IFCs with respect to *B* values for the magnetoexciton $|n{=}41\rangle \rightarrow |n'{=}41\rangle$ in free-suspended monolayer $WTe_2$ ($\hat{\sigma}_b = 0$, $\varepsilon_2 = \varepsilon_3 = 1$). Inset in (b) shows the partially magnified IFCs for clarity. IFCs for monolayer $WTe_2$ deposited on different negative-$\varepsilon_3$ substrates ($\hat{\sigma}_b$=0, $d$=0) at the transitions $|n{=}6\rangle \rightarrow |n'{=}6\rangle$ (c) and $|n{=}41\rangle \rightarrow |n'{=}41\rangle$ (e). IFCs versus the distance *d* between $WTe_2$ and the substrate for (d) $|n{=}6\rangle \rightarrow |n'{=}6\rangle$ and (f) $|n{=}41\rangle \rightarrow |n'{=}41\rangle$. In (d)(f), the factors are $\hat{\sigma}_b$=0, $\varepsilon_2$=1, $\varepsilon_3$=−5, $\lambda_6$=1242.1 nm and $\lambda_{41}$=1102.5 nm.

## III. RESULTS AND DISCUSSIONS

We first consider the simplest condition that the upper $WTe_2$ is suspended in a free space ($\varepsilon_3$=$\varepsilon_2$=1 and $\hat{\sigma}_b$=0). Fig. 1(b) presents partially magnified $\sigma_{xx}$, $\sigma_{yy}$ and $\sigma_{yx}$ spectra of monolayer $WTe_2$. Their overall spectra ranging the $\hbar\omega$ from 0.98 to 1.20 eV are given in section 1 in Ref. [50]. It is found that all the conductivity spectra exhibit SdH-like oscillations, and a three-peak structure is well resolved in each oscillation period in the upper subgraph in Fig. 1(b). By matching with LLs in Fig. S1 [50], these three peaks from left to right correspond to interband transitions when the LL index changes $\delta n = n' - n = -2$, 0, and +2. Oscillation periods at *n*=6 and 41 are marked by two dotted-line boxes in Fig. 1(b), and interband transition rules are sketched in Fig. 1(c).

Figs. 1(e)(f) show IFCs of HMEPs $\delta n$=0, ±2 when the LL index *n*=6 and 41. In them, closed elliptic IFCs originate from *s*-polarized wave, which are insensitive to the variation of LL index. For *p*-polarized mode, different $\delta n$ transitions give different hyperbolic IFCs. Especially, two almost parallel isofrequency curves are observed at $\delta n$=0 in Fig. 1(e), yielding that the divergence angle $\theta$ (between group velocity $\boldsymbol{v_g}$ and $k_x$ axis) is nearly zero at $|k_y|$>$10k_0$. This strongly directional propagation constitutes a clear signature of polariton canalization. According to Fig. 1(b), the magnetoexciton $|n{=}6\rangle \rightarrow |n'{=}6\rangle$ has the largest numerical differences between $\sigma_{xx}$, $\sigma_{yy}$, and $\sigma_{yx}$, indicating the strongest anisotropy induced via SdH effect. For interband transitions $\delta n$=−2 and +2 (*n*=6), the conductivity difference from the former is larger than that from the latter (detailed comparison is given in Fig. S5) [50]. Thus, the canalization of $|n{=}6\rangle \rightarrow |n'{=}4\rangle$ is better than that of $|n{=}6\rangle \rightarrow |n'{=}8\rangle$. In 1st quadrant, their divergence angles at $k_y$=$30k_0$ are $\theta$=1.8° and 23.8°, respectively.

The evolution of IFCs with increasing LL index *n* ($\delta n$=0) from 1 to 61 is uncovered in Figs. 2(a) and S8 (Fig. S8 is a detailed version). IFCs of *p*-polarized wave are found to possess two-fold open hyperbolas at *n*<41 while they transit to one-sheet open ones as LL index increases to be *n*≥41. Besides, because the anisotropy of conductivity tensor gradually becomes weaker (*cf.* conductivity components in Fig. S5), the $\theta$ value increases from 0.7° to 20.1° as LL index *n* increases from 1 to 61 (see Fig. S9), implying that high-energy-level HMEPs are relatively difficult to be canalized in $WTe_2$. Comparing with $\theta$≈0° in Fig. 1(e), the computed $\theta$ value marked in Fig. 1(f) for $|n{=}41\rangle \rightarrow |n'{=}41\rangle$ is 11.5°. Even so, the canalization (or IFC topology) of high-LL HMEPs can be improved (or modulated) by increasing the *B*. As demonstrated in Figs. 2(b), one-sheet hyperbola of $|n{=}41\rangle \rightarrow |n'{=}41\rangle$ gradually transfers into two-fold ones as *B* increases to 25 T [accompanied with broadening of LL spacings, as illustrated in Fig. 1(c)], both its topology and curve shape at *B*=25 T become similar to that of $|n=6\rangle\rightarrow|n'=6\rangle$ in Figs. 1(e) and 2(a) at $B$=10 T. This similarity is also verified by our simulated field distributions [*cf.* Figs. S7(b)(h) and S20(c)(f)]. Note that the value range of $k_x$ is far smaller than that of $k_y$ in Figs. 1(e)(f) to better distinguish IFC topologies, visually, the included angle $\theta$ at nonequal-ratio $k_x$ and $k_y$ scales may deviate from the actual angle between the group velocity $\boldsymbol{v_g}$ and the normal of optical axis. Hyperbolic IFCs at equal ratios of $k_x$ and $k_y$ axes along with their field distributions are given in Figs. S6-S7 where canalization phenomena are better visualized.

Canalized HMEPs are further delineated by azimuthal dispersions of surface waves. As shown in Fig. 1(d), real part Re($q$) of surface polariton wavevector $q$ at $\phi$=90° is more than 10 times larger than that at $\phi$=0°, further confirming the canalization direction along $y$ direction. Combining with the imaginary part Im($q$) in Fig. S10, the group velocity and lifetime of HMEPs along $y$ direction have been computed by $v_g=|\partial\omega/\partial q|$ and $\tau=1/[v_g\mathrm{Im}(q)]$, respectively [1,36]. Results show that both $v_g$ and $\tau$ are fluctuant as the LL index changes [see Fig. 5(a)]. For HMEPs $|n=6\rangle\rightarrow|n'=6\rangle$ and $|n=41\rangle\rightarrow|n'=41\rangle$, their group velocities (lifetimes) are $2.28\times10^{-5}c$ and $6.20\times10^{-7}c$ (244 μs and 444 μs), respectively. Compared with the group velocity ($\sim10^{-3}c$) and lifetimes (several picoseconds) of canalized phonon polaritons [1,5,36,37], HMEPs relax in extremely slow velocities and exhibit super-long relaxation times.

To inspect the influence of dielectric environment on topology and canalization of HMEPs with $p$-polarization excitation, Figs. 2(c) and (e) show IFCs of $|n=6\rangle\rightarrow|n'=6\rangle$ and $|n=41\rangle\rightarrow|n'=41\rangle$ in $WTe_2$ deposited on different negative-$\varepsilon_3$ substrates, respectively. Impacts of positive-$\varepsilon_3$ substrates on IFCs are shown in Figs. S11(a)(b). It is found that increasing positive $\varepsilon_3$ values has no impact on IFC topologies of both transitions. However, the substrate with $\varepsilon_3=-1$ induces both hyperbolic IFCs to be transformed to a kind of curve similar to witch of Agnesi [52]. Further reducing the $\varepsilon_3$ to be $\leq-3$, IFCs for $|n=41\rangle\rightarrow|n'=41\rangle$ are relatively simple, which are transformed to two-fold hyperbolas [see Fig. 2(e)]. By contrast, IFCs for $|n=6\rangle\rightarrow|n'=6\rangle$ are complicated, which transfer into two-fold hyperbolas accompanied with two unclosed arcs near the origin of coordinate axes at $\varepsilon_3=-3$ and $-5$, as plotted by green and blue dashed lines in Fig. 2(c). As the $\varepsilon_3$ further decreases to $-7$ and $-9$, IFCs are transformed to one-sheet hyperbolas in Fig. 2(c). For $s$-polarization, the IFC topology (closed ellipsoid) is not shown here because it is little affected by varied $\varepsilon_3$ values.

The thickness $t$ of monolayer $WTe_2$ is only ~7.9 Å such that magnetoexcitons are confined in a 2D plane and the Coulomb field permeates both 2D material and surrounding media [53,54]. Thus, evolutions of IFCs with the $\varepsilon_3$ could be associated with substrate-incurred non-local dielectric screening effect on magnetoexcitons. The effective radius of magnetoexciton depends on LL index, and can be calculated by $a_n=a_B[3n(n\text{-}1)+1]$ with the three-dimensional Bohr radius $a_B$ [54-58]. The effective radius of magnetoexciton $|n=6\rangle\rightarrow|n'=6\rangle$ ($|n=41\rangle\rightarrow|n'=41\rangle$) is estimated to be $a_{n=6}$=6.8 Å ($a_{n=41}$=26.1 Å) [50]. Thereby, both spatial separations ($2a_n$) of Coulomb-bound electron-hole pairs for $n$=6 and 41 are larger than the thickness $t$. A portion of the electric field permeate into top and bottom media of monolayer $WTe_2$, as sketched in Fig. 3(a). By using the method of image charges [50], Coulomb potential distributions of magnetoexcitons, which depend on negative and positive $\varepsilon_3$ values, are simulated, as shown in Figs. 3(b)(c) and S15, respectively.

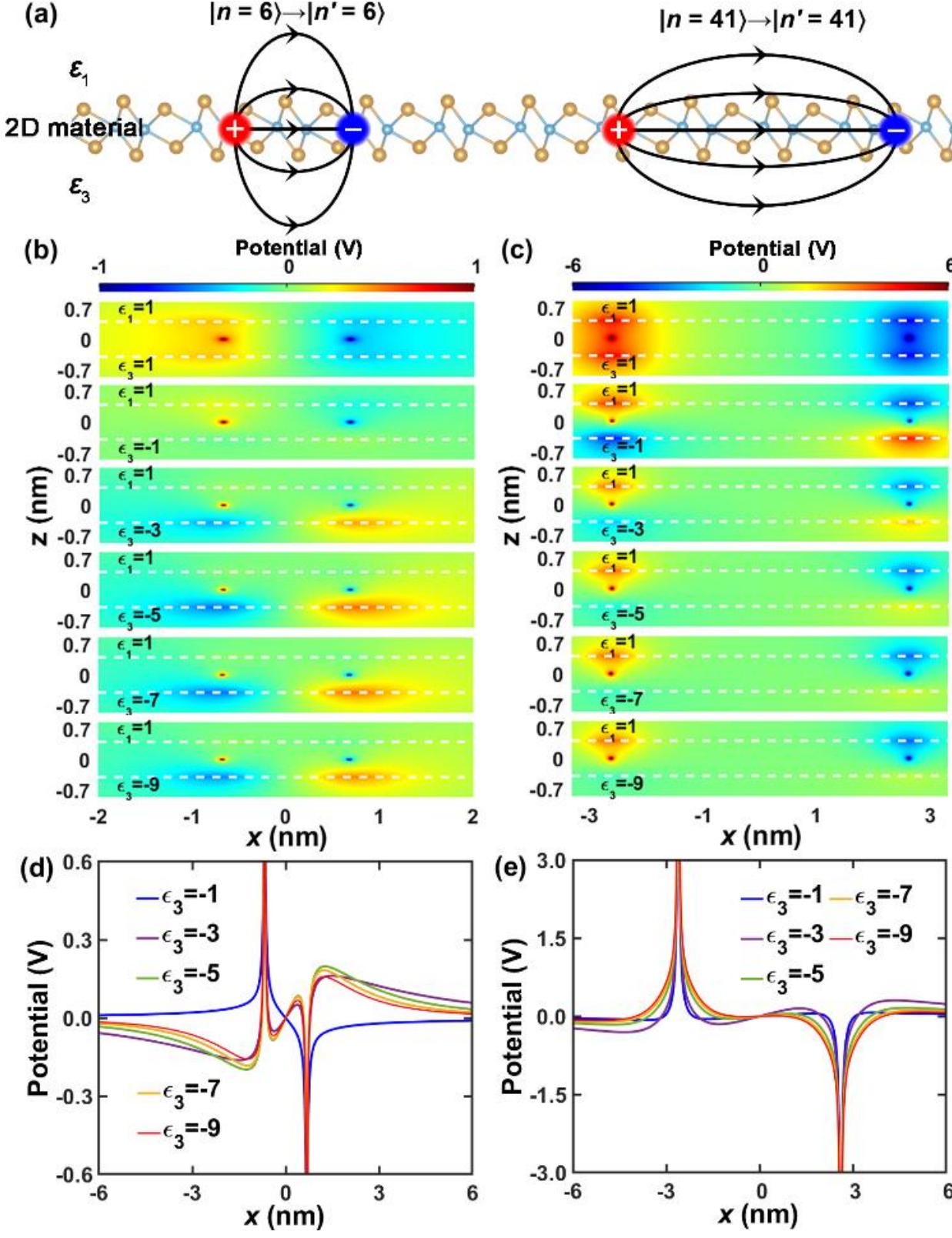

FIG. 3. (a) Schematic diagram of externally induced dielectric screening effects on magnetoexcitons $|n=6\rangle\rightarrow|n'=6\rangle$ and $|n=41\rangle\rightarrow|n'=41\rangle$. (b)(c) Coulomb potential distributions with a positive and a negative unit charge in monolayer $WTe_2$ deposited on a negative-$\varepsilon_3$ substrate. Interfaces between $WTe_2$ and the air (or substrate) are denoted by white dashed lines. (d)(e) Distribution curves of Coulomb potentials at the $z$ = 0 line in (b)(c). (b)(d) and (c)(e) are for $|n=6\rangle\rightarrow|n'=6\rangle$ and $|n=41\rangle\rightarrow|n'=41\rangle$, respectively. Note that $x$-axis range of Coulomb potentials is [-6 6] in practical calculations. To clearly present the potential distributions in the vicinity of unit charges, $x$-axis ranges in (b) and (c) are [-2 2] and [-3.3 3.3], respectively.

We see from Fig. S15 that positive $\varepsilon_3$ does not change the sign of Coulomb potentials, and potential distributions are mirror symmetry about the axis of $x$=0. That is, potentials are positive (or negative) at the side of positive (or negative) unit charge. Nevertheless, negative $\varepsilon_3$ breaks this mirror symmetry, and the sign of Coulomb potentials distributing at the space $z$<0 is opposite to that of unit charges. As demonstrated in Figs. 3(b)(c), the positive (or negative) unit charge locates at $x=-a_n$ (or $+a_n$), but potentials are negative (or positive) at 3rd (or 4th) quadrant. Distribution curves of potentials at $z$=0 line, which are extracted from Figs. 3(b)(c) and are given in Figs. 3(d)(e), show that $\varepsilon_3=-1$ leads potentials to be close to zero except for the extreme positions of unit charges. When $\varepsilon_3 \leq -3$, the Coulomb potential of magnetoexciton $|n=6\rangle \rightarrow |n'=6\rangle$ is confined at the interface between $WTe_2$ layer and air [see Fig. 3(b)] while that of magnetoexciton $|n=41\rangle \rightarrow |n'=41\rangle$ is confined at the interface between $WTe_2$ layer and the substrate [see Fig. 3(c)]. Moreover, Coulomb potentials near magnetoexciton $|n=41\rangle \rightarrow |n'=41\rangle$ are monotonically changed with negative $\varepsilon_3$. For example, potentials at the interface between $WTe_2$ layer and air are gradually strengthened with reducing the $\varepsilon_3$ from −3 to −9. In mark contrast, variations of Coulomb potentials near magnetoexciton $|n=6\rangle \rightarrow |n'=6\rangle$ are nonmonotonic. For instance, potentials at the interface between $WTe_2$ layer and the substrate become the strongest with reducing the $\varepsilon_3$ to −5. Thereafter, they are diminished with a reduction of the $\varepsilon_3$. Hence, IFCs of $|n=6\rangle \rightarrow |n'=6\rangle$ exhibit richer optical topologies than that of $|n=41\rangle \rightarrow |n'=41\rangle$ at modulating the $\varepsilon_3$ in a negative-value range.

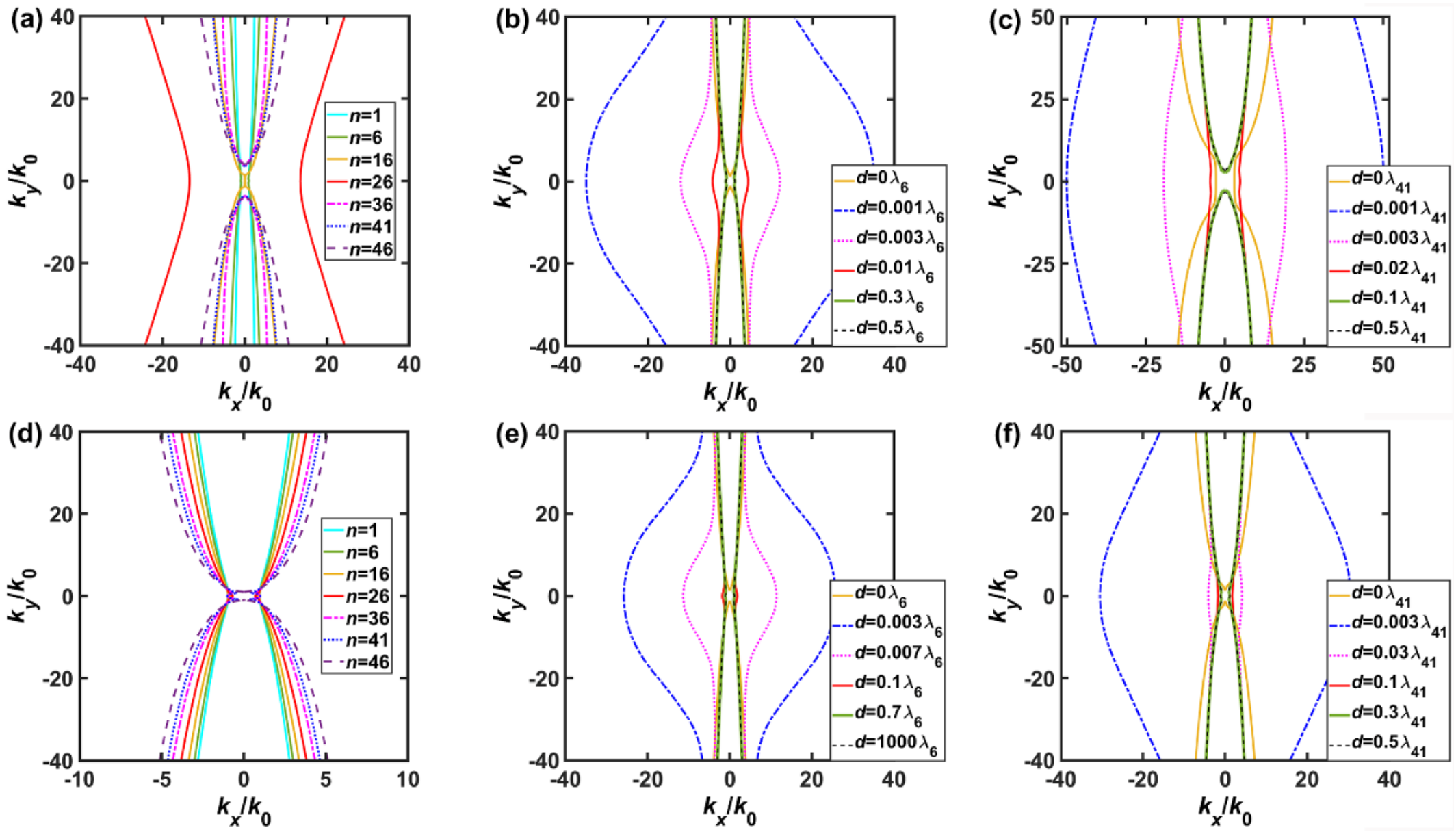

FIG. 4. IFCs of free-suspended monolayer $MoS_2$ (a) and phosphorene (d) versus LL index $n$ at $\delta n = 0$, $B = 10$ T, and $p$-polarization excitation. Influences of the distance $d$ between monolayer materials and the substrate on IFCs for (b)(e) $|n = 6\rangle \rightarrow |n' = 6\rangle$ and (c)(f) $|n = 41\rangle \rightarrow |n' = 41\rangle$ with factors $\varepsilon_1 = \varepsilon_2 = 1$, $\varepsilon_3 = -5$. (b)(c) are for the case of $MoS_2$ with $\lambda_6 = 1742.3$ nm and $\lambda_{41} = 1231.0$ nm. (e)(f) are for the case of phosphorene with $\lambda_6 = 791.3$ nm and $\lambda_{41} = 669.7$ nm.

Even if positive and negative substrates have different impacts on optical topologies of IFCs, they have a similar influence on the openings of hyperbolic IFCs. At positions away from magnetoexcitons, e.g., $|x|$>2.5 nm for $|n=6\rangle \rightarrow |n'=6\rangle$ and $|x|$>4.0 nm for $|n=41\rangle \rightarrow |n'=41\rangle$ at $z$=0 line [see Figs. 3(d)(e) and S16(a)(b)], Coulomb potentials are mainly diminished as the $|\varepsilon_3|$ value increases. Meanwhile, all the openings of hyperbolic IFCs in Figs. 2(c)(e) and S11(a)(b) become larger and larger as the $|\varepsilon_3|$ value increases. This means that a larger $|\varepsilon_3|$ induces to a stronger screening effect on Coulomb potentials away from magnetoexcitons, which is detrimental to the canalizations of HMEPs.

If there is a thin free-space interlayer between monolayer $WTe_2$ and the substrate (e.g., $\varepsilon_2$=1, $\varepsilon_3$=−5, and $d=0.001\lambda_n$, $\lambda_n$ is the excitation wavelength), hyperbolic IFCs in Figs. 2(c)(e) will morph into the expansive witch of Agnesi in Figs. 2(d)(f). The maximum $|k_x/k_0|$ of witch of Agnesi can reach to 49.0 and 61.8 for $|n=6\rangle \rightarrow |n'=6\rangle$ and $|n=41\rangle \rightarrow |n'=41\rangle$, respectively. For clarity, $k_x/k_0$ values for the case of $d=0.001\lambda_n$ in Figs. 2(d) and (f) are multiply by 0.35 and 0.5, respectively. In Fig. 2(d), the IFC displays nearly twin vertical lines aligned with $k_y$ axis when the $d$ increases to $0.03\lambda_6$, demonstrating a robust polariton canalization. When the $d$ is large enough, the screening effect of substrate $\varepsilon_3$ on magnetoexcitons will disappear, inducing IFCs to revert to hyperbolic topologies of free-suspended $WTe_2$ [*cf.* green curves in Figs. 2(d)(f) and fiery-red curves in Figs. 1(e)(f)].

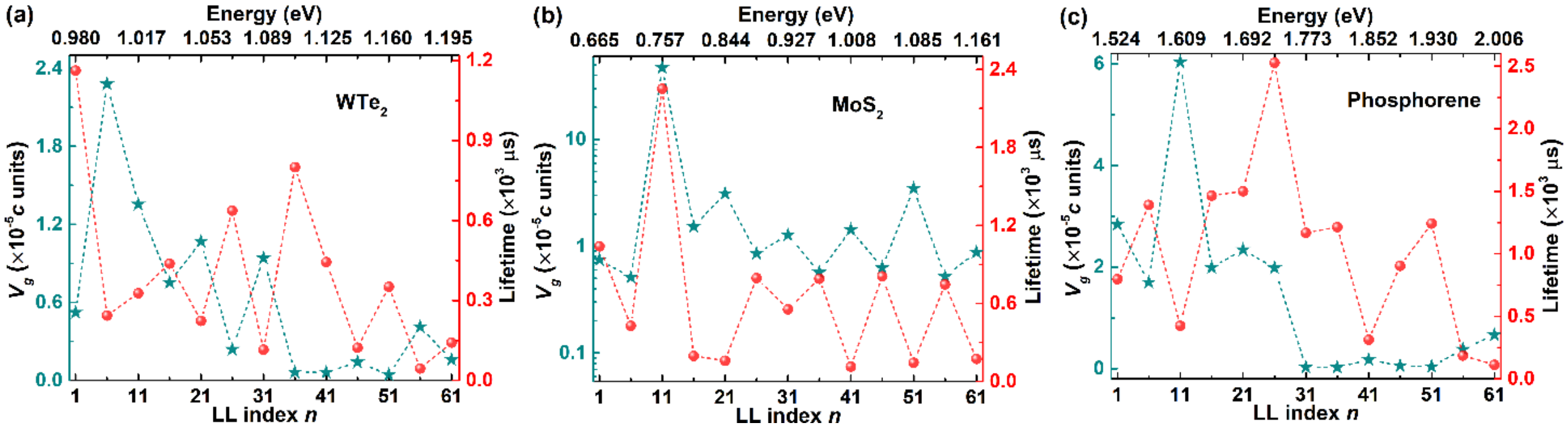

FIG. 5. For HMEPs along zigzag orientation in free-suspended monolayer (a) $WTe_2$, (b) $MoS_2$, and (c) phosphorene, group velocities $v_g$ (see pentacle symbols) and lifetimes $\tau$ (see spherical symbols) as a function of the LL index $n$ of interband transitions $\delta n = n'-n = 0$. Symbols are the calculated values while dashed lines are just drawn as a guide to the eye.

As illustrated in Fig. 3(a), the spatial separation $2a_n$ between positive and negative charges for $|n=6\rangle\rightarrow|n'=6\rangle$ is far shorter than that for $|n=41\rangle\rightarrow|n'=41\rangle$. Thus, the electric field from the former spreads to a longer distance along the vertical direction. Consequently, a lager $d$ value is necessary for $|n=6\rangle\rightarrow|n'=6\rangle$ to eliminate the interaction between magnetoexcitons and the substrate. Based on this, screening lengths of HMEPs $|n=6\rangle\rightarrow|n'=6\rangle$ and $|n=41\rangle\rightarrow|n'=41\rangle$ are estimated, which are $0.3\lambda_6$=372.6 nm and $0.1\lambda_{41}$=110.3 nm, respectively [see Figs. 2(d)(f)].

We also inspect vdW crystals of 1T′-$MoS_2$ and phosphorene monolayers. Figs. S22-S26 (or S29-S32) show LLs and complex optical conductivity spectra of monolayer $MoS_2$ (or phosphorene). SdH-like quantum oscillations well reappear in the conductivity spectra. LL index-dependent hyperbolic IFCs and canalizations can also be found from free-suspended $MoS_2$ in Fig. 4(a) (its detailed version is Fig. S27) and phosphorene in Figs. 4(d) (its detailed version is Fig. S33). The calculated group velocities of HMEPs in Figs. 5(b)(c) show that they are also at the order of magnitudes ~$10^{-5}c$, and the HMEP $|n=26\rangle\rightarrow|n'=26\rangle$ in monolayer phosphorene has the longest lifetime which can be up to 2.5 ms. Differing from $MoS_2$ and $WTe_2$, Figs. 4(d) and S33 show that all $p$-polarization excited IFCs in phosphorene exhibit highly canalized hyperbolas and their divergence angles $\theta$ only increase from 2.2° to 4.8° with ranging LL index $n$ ($\delta n = 0$) from 1 to 61 [see Fig. S34]. The strongest canalization may originate from the most robust anisotropy induced via SdH effect in phosphorene. As shown in Fig. S30, the overall trends in amplitudes of real and imaginary parts of $\sigma_{xx}$ (or $\sigma_{yy}$) are gradually decreased (or increased) with increasing photon energy. This contrary variation trend between $\sigma_{xx}$ and $\sigma_{yy}$ makes phosphorene have a strong anisotropy not only at low LL transitions but also at high LL transitions.

Figs. 4(b)(c) [or 4(e)(f)] present the influence of distance $d$ between monolayer $MoS_2$ (or phosphorene) and the substrate ($\varepsilon_3$=−5) on IFCs for HMEPs $|n=6\rangle\rightarrow|n'=6\rangle$ and $|n=41\rangle\rightarrow|n'=41\rangle$. It is found that curve shapes of IFCs in Figs. 4(b)(c) [or 4(e)(f)] tend to be stable when the distance $d$ increases to be $\geq 0.3\lambda_6$ and $\geq 0.1\lambda_{41}$ [or ($\geq 0.7\lambda_6$ and $\geq 0.3\lambda_{41}$)] for $|n=6\rangle\rightarrow|n'=6\rangle$ and $|n=41\rangle\rightarrow|n'=41\rangle$, respectively. Thereby, the estimated screening lengths of HMEPs $|n=6\rangle\rightarrow|n'=6\rangle$ and $|n=41\rangle\rightarrow|n'=41\rangle$ in monolayer $MoS_2$ (phosphorene) are about $0.3\lambda_6$=522.7 nm and $0.1\lambda_{41}$=123.1 nm ($0.7\lambda_6$=553.9 nm and $0.3\lambda_{41}$=200.9 nm), respectively. It should be emphasized here that optical conductivities used for dispersion Eqs. (6) and (7) does not take into nonlocal corrections [21,59,60]. In practice, nonlocal corrections are expected to significantly reduce HMEPs' quantum confinement [21,59,60], leading factual $|k_x/k_0|$ values of witch of Agnesi in Figs. 2 and 4 to be smaller than our calculated values.

The canalization direction and topological state of HMEPs can be further programmed via twisting bilayer 2D materials. Fig. 6 shows IFCs of sandwich structure consisting of two $WTe_2$ monolayers with an interlayer between them. At $d$=$0\lambda_n$ and $\Delta\phi$=0°, IFCs for transitions $|n=6\rangle\rightarrow|n'=6\rangle$ and $|n=41\rangle\rightarrow|n'=41\rangle$ exhibit one-sheet and two-fold hyperbolas, respectively. Their divergence angles marked in Figs. 6(a)(b) are $\theta\approx0°$ and 6.1°, respectively. When an interlayer $\varepsilon_2$=−1 is introduced, hyperbolic IFCs of $|n=6\rangle\rightarrow|n'=6\rangle$ and $|n=41\rangle\rightarrow|n'=41\rangle$ morph into one-sheet and two-fold pincerlike IFCs, respectively. The intersections between two "pincer's handles" gradually shift toward $k_y/k_0$=0, and the openings of IFCs gradually get larger as the $|\varepsilon_2|$ increases (the condition of positive-$\varepsilon_2$ interlayer is given in Fig. S12). As the $\varepsilon_2$ reduces to −5, two-fold pincerlike IFCs in Fig. 6(b) are transformed to one-sheet ones.

If there is a nonzero twisted angle $\Delta\phi$ between upper and bottom $WTe_2$ layers, following recently proposed hyperbolic shear phonon polaritons [62-66], pincerlike shear magnetoexciton polaritons are obtained, as shown in Figs. 6(c)(d). Two "pincer's handles" rotate in opposite direction as the $\Delta\phi$ increases, and they will be completely opened up when the $\Delta\phi$ is large enough. The canalization direction along the optical axis of IFC features a clockwise rotation with increasing the twisted angle. Canalizations of pincerlike HMEPs in Fig. 6 are better visualized at equal ratios of $k_x$ and $k_y$ axes in Fig. S35. Besides, divergence angles marked in Figs. 6(c)(d) for $|n=6\rangle\rightarrow|n'=6\rangle$ and $|n=41\rangle\rightarrow|n'=41\rangle$ are $\theta$=7.4° and 10.7°, respectively, further demonstrating that pincerlike shear magnetoexciton polaritons are endowed with extreme anisotropic propagations.

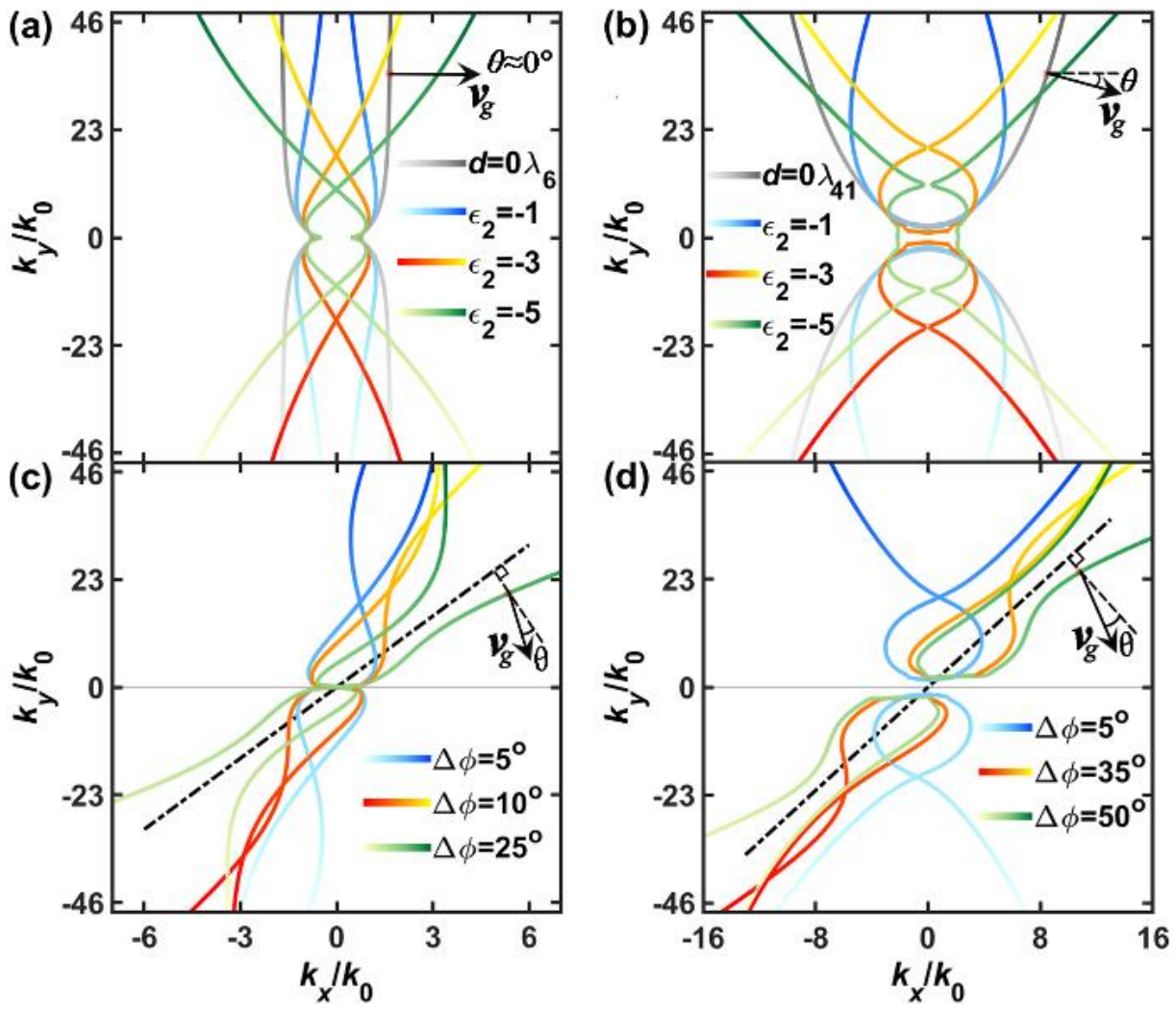

FIG. 6. For the sandwich structure consisting of two $WTe_2$ monolayers with an interlayer between them, (a)(b) IFCs of sandwich structure at $d$=0$\lambda_n$ and $d$=0.01$\lambda_n$ ($\varepsilon_2$=−1, −3, and −5). The twisted angle $\Delta\phi$=0°. (c)(d) IFCs of sandwich structure versus the $\Delta\phi$ at $d$=0.01$\lambda_n$ and $\varepsilon_2$=−3. The divergence angle $\theta$ is the included angle between the group velocity $\boldsymbol{v_g}$ and the normal of optical axis. Optical axes at $\Delta\phi$=25° and 50° are marked by dot-dashed lines. (a)(c) and (b)(d) are for the transitions $|n=6\rangle\rightarrow|n'=6\rangle$ and $|n=41\rangle\rightarrow|n'=41\rangle$, respectively. The other factors are $\varepsilon_1$=$\varepsilon_3$=1, $\lambda_6$=1242.1 nm, and $\lambda_{41}$=1102.5 nm.

## IV. CONCLUSIONS

To conclude, our work predicts canalized HMEPs which derive from SdH effect in natural $WTe_2$, $MoS_2$, and phosphorene monolayers. Because bandgaps of these three materials are located in the visible to near-infrared frequencies, canalized HMEPs occur in the visible to near-infrared regimes. Moreover, SdH effect is theoretically suitable for monocrystals of any substances under the conditions of ultralow temperature and intense magnetic field, hence canalized magnetoexciton polaritons could be achieved in all vdW crystals, and be tuned in any desired frequency ranges via magnetizing appropriate bandgap semiconductors. In addition, our predicted HMEPs feature ultralow group velocity (~$10^{-5}c$), super-long lifetime (hundreds of microseconds), and multiple IFC topologies including witch-of-Agnesi, one-sheet, two-fold, and shear pincerlike IFCs. These exotic IFCs provide more approaches to tailor wavefronts in richer shapes. LLs and twisted angles can be used to program the canalizations, and a new method based on IFCs of HMEPs is proposed to evaluate screening lengths of magnetoexcitons formed via different LL transitions. Our findings open new vistas for controlling and steering the energy flux and topology of hyperbolic polaritons at will at the nanoscale.

## ACKNOWLEDGMENTS

G. Jia acknowledges the financial support by the National Natural Science Foundation of China (Grant No. 11804251). C. Qiu acknowledges support from the Ministry of Education in Singapore (Grant Nos. A-8002152-00-00 and A-8002458-00-00). The authors are grateful to Dr. Xuezhi Ma from A*STAR Quantum Innovation Centre (Q. InC) in Singapore, who provided support on calculating near-field distributions.

## DATA AVAILABILITY

The data that support the findings of this article are not publicly available. The data are available from the authors upon reasonable request.

Corresponding authors:
*gyjia87@163.com; †chengwei.qiu@nus.edu.sg